\documentclass{aastex61}
\usepackage{graphicx}
\usepackage{times}
\begin{document}

% The following seven commands are intended for editorial usage and should be ignored by
% the author(s).
%\Pagespan{789}{}% Document's page range. 
% If second parameter is left empty, the last page is computed automatically.
%\Yearpublication{2006}%
%\Yearsubmission{2005}%
%\Month{11}%   
%\Volume{999}%  
%\Issue{88}% 
% \DOI{This.is/not.aDOI}% 

\title{
BVRI Photometry of SN 2016coj in NGC 4125
}

\author{Michael W. Richmond}
\shorttitle{SN 2016coj in NGC 4125 }
\shortauthors{M. W. Richmond and B. Vietje }
\affil{
Physics Department, Rochester Institute of Technology,
84 Lomb Memorial Drive, Rochester, NY, 14623 USA
}
\email{mwrsps@rit.edu}
\author{Brad Vietje}
\affil{
Northeast Kingdom Astronomy Foundation,
PO Box 173, Peacham, VT  05862 
}
\email{brad@nkaf.org}

\keywords{ AAVSO keywords = supernovae: individual (SN 2016coj) }
\keywords{ ADS keywords = supernovae: individual (SN 2016coj) }

\begin{abstract}

We present BVRI photometry of supernova (SN) 2016coj in NGC 4125
from $9$ days before to $57$ days after its $B$-band maximum light.
Our light curves and color curves suggest that 
this event belongs to the ``normal'' class of 
type Ia SNe,
with a decline rate parameter
$\Delta m_{15}(B) = 1.32 \pm 0.10$,
and that it suffers little extinction.
Adopting a distance modulus to its host galaxy
of 
$(m - M) = 31.89$ mag,
we compute extinction-corrected
peak absolute magnitudes of
$M_B = -19.01$,
$M_V = -19.05$,
$M_R = -19.03$,
and 
$M_I = -18.79$.
The explosion occurred close enough to the
nucleus of NGC 4125 to hinder 
the measurement of its brightness.
We describe our methods to reduce the 
effect of such host-galaxy
contamination,
but it is clear that our latest values suffer from systematic bias.

\end{abstract}

%\maketitle

\section{Introduction}

Supernova (SNe) of type Ia are thought to 
originate in close binary systems,
consisting of either a single white dwarf and
a main-sequence companion, 
or two white dwarfs.
When one white dwarf accretes enough material
to exceed the Chandrasekhar limit
\citep{Chan1931},
either by long-term transfer from a main sequence companion,
or by a violent merger with another white dwarf,
a runaway thermonuclear reaction propagates 
through it, disrupting the entire white dwarf,
heating the ejecta to hundreds of thousands of degrees
and blowing it out into space at thousands of
kilometers per second.
The expanding cloud of hot gas radiates energy
for several months, 
reaching absolute magnitudes in the optical of 
order -18 to -20.
Many (but not all) type Ia SNe exhibit similar properties,
with a correlation between the shape of the light curve
and the absolute magnitude at peak
\citep{Phil1993}.
When events are observed in sufficient detail,
one can use the shape of the light curve to compute
the absolute magnitude
(\citet{Prie2006}; \citet{Guy2005}),
and so use these SNe
as ``standard-izable candles'' to determine distances.

Supernova 2016coj in the galaxy NGC 4125,
a peculiar elliptical of class E6 \citep{Deva1991},
was discovered by the Lick Observatory Supernova 
Search 
(\citealt{Fili2001}; \citealt{Leam2011})
on UT 2016 May 28
\citep{Zhen2016}
and quickly identified as a type Ia explosion.
Since its host galaxy is relatively close
to our Milky Way,
at a redshift of only $z = 0.004523$
according to NED,\footnote
{NASA Extragalactic Database, see
  {\url{https://ned.ipac.caltech.edu}} }
this event promised to provide a wealth of
high-precision information.
However, since the supernova occurred not far from
the galaxy's nucleus, disentangling its light from
that of the surrounding stars turns out to be 
a difficult task.

In this paper, we describe photometry of SN 2016coj
in the BVRI passbands acquired at two locations,
starting on UT 2016 May 30 and ending UT 2016 Aug 4,
an interval of 66 days.
Section 2 describes our observational methods,
the cleaning of the raw CCD images,
and the techniques we used to extract instrumental magnitudes.
We explain our photometric calibration of the
raw measurements onto the standard Johnson-Cousins
system in Section 3.
The light curves and color curves of the 
event are shown in Section 4;
we comment briefly on their properties and
the effect of extinction along the line of sight.
We present our conclusions in Section 5.
In an appendix, we discuss the difficulties of
measuring the light of a point source immersed
in a non-uniform background,
and use simple simulations to estimate the nature
of systematic biases that appear in our data.

\section{Observations}

We present herein data acquired at
the RIT Observatory, near Rochester, New York,
and
at the Northern Skies Observatory (NSO),
in Peacham, Vermont.
We will describe below the procedures by which we
acquired and reduced the images from 
each observatory in turn.

The RIT Observatory is located on the southeastern
corner of the Rochester Institute of Technology campus,
at 
longitude 77:39:53 West, latitude +43:04:33 North,
and an altitude of 168 meters.
Our Meade LX200 f/10 30-cm telescope provides a 
plate scale of 
$1{\rlap.}^{''}38$
per pixel at the focus of our SBIG ST-9 camera,
which has BVRI filters built to the Bessell prescription.
When observing SN 2016coj,
we acquired a series of 5 to 20 short exposures
(exposure times 30 seconds each up to 2016 July 19 = JD 2457588,
120 seconds each after that date),
discarding those with trailing or extinction by clouds.
We acquired dark and flatfield images each night, creating master
frames from the median of 10 individual images.
Flatfields were based on images of the twilight sky,
with the exception of UT June 13, when bad conditions forced us to
use dome flats.
After subtracting the master dark from each target frame
and dividing it by the normalized master flatfield,
we examined each resulting ``clean'' image by eye,
discarding those with poor quality.

\begin{figure}
\plotone{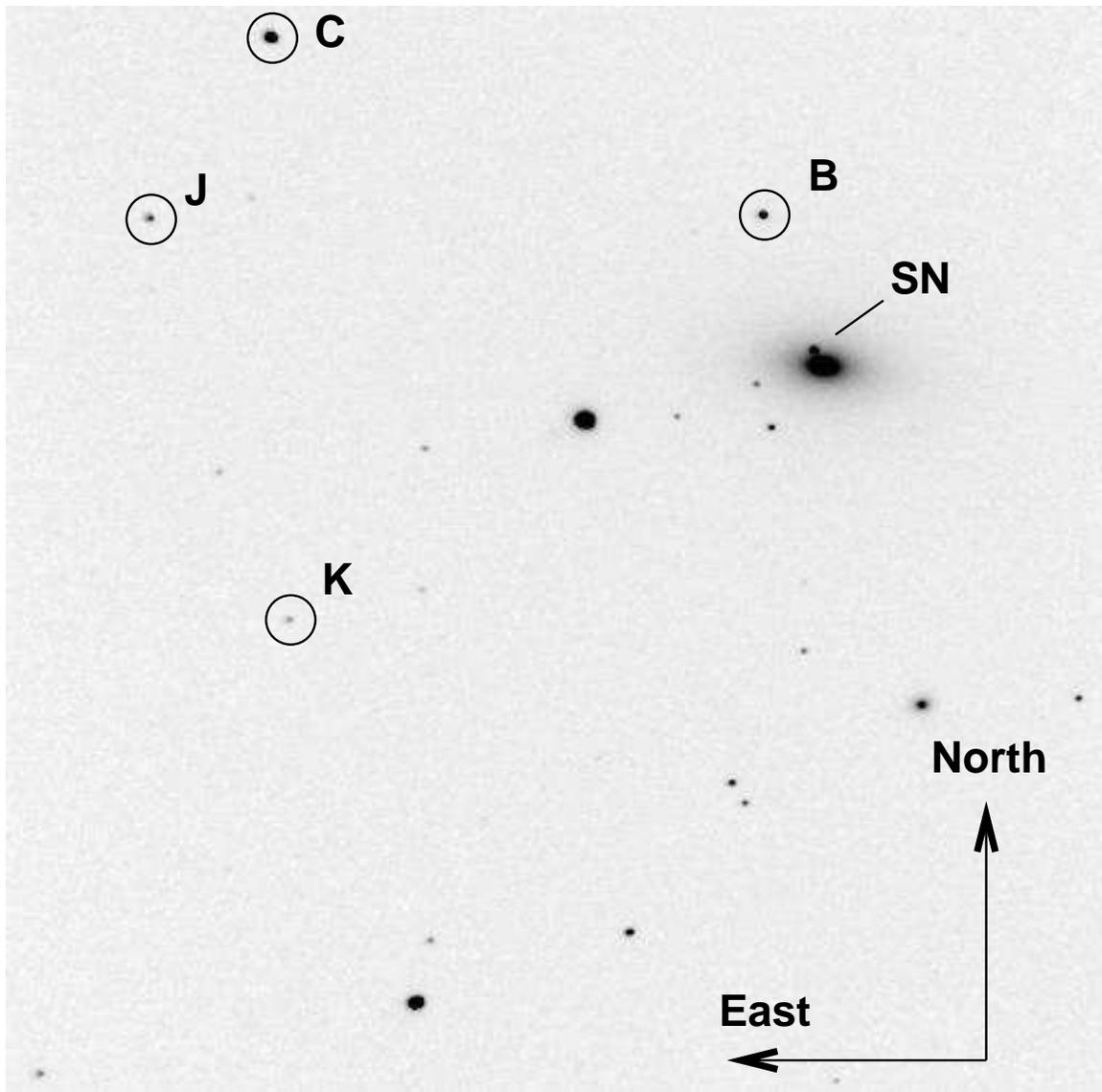}
\caption{A R-band composite (16 images of 30 seconds each)
         of SN 2016coj from RIT,
         showing stars used to calibrate measurements.
         North is up, East to the left.
         The field of view is roughly 12 by 12 arcminutes.
         \label{fig:chartritlabel} }
\end{figure}

Before extracting instrumental magnitudes, 
we combined all the images in a particular passband
using a median technique,
in order to increase the signal-to-noise ratio
and eliminate cosmic rays.
Figure \ref{fig:chartritlabel}
shows an example of such a combined image, 
with labels indicating stars used for calibration.
The Point Spread Function (PSF) of these combined images had a typical 
Full Width at Half Maximum (FWHM) 
of 
$3{\rlap.}^{''}5$ 
to
$4{\rlap.}^{''}1$.

Since the supernova lies only about 12 arcseconds
from the nucleus of its host galaxy
\citep{Zhen2016},
simple aperture photometry will yield
poor results.
A systematic error can appear in such measurements
due to imperfect background subtraction;
the size of the error will grow as the supernova fades.
A standard technique in such cases is to match
each target image to a template of the same galaxy
taken some time before or after the event,
in which the supernova does not appear,
and then to subtract the template from the target image.
However, since we lacked template images, 
we adopted a technique which does not require them;
the drawback is that its results are less accurate,
and can still suffer from systematic effects.
See the Appendix for a detailed explanation.

\begin{figure}
\plotone{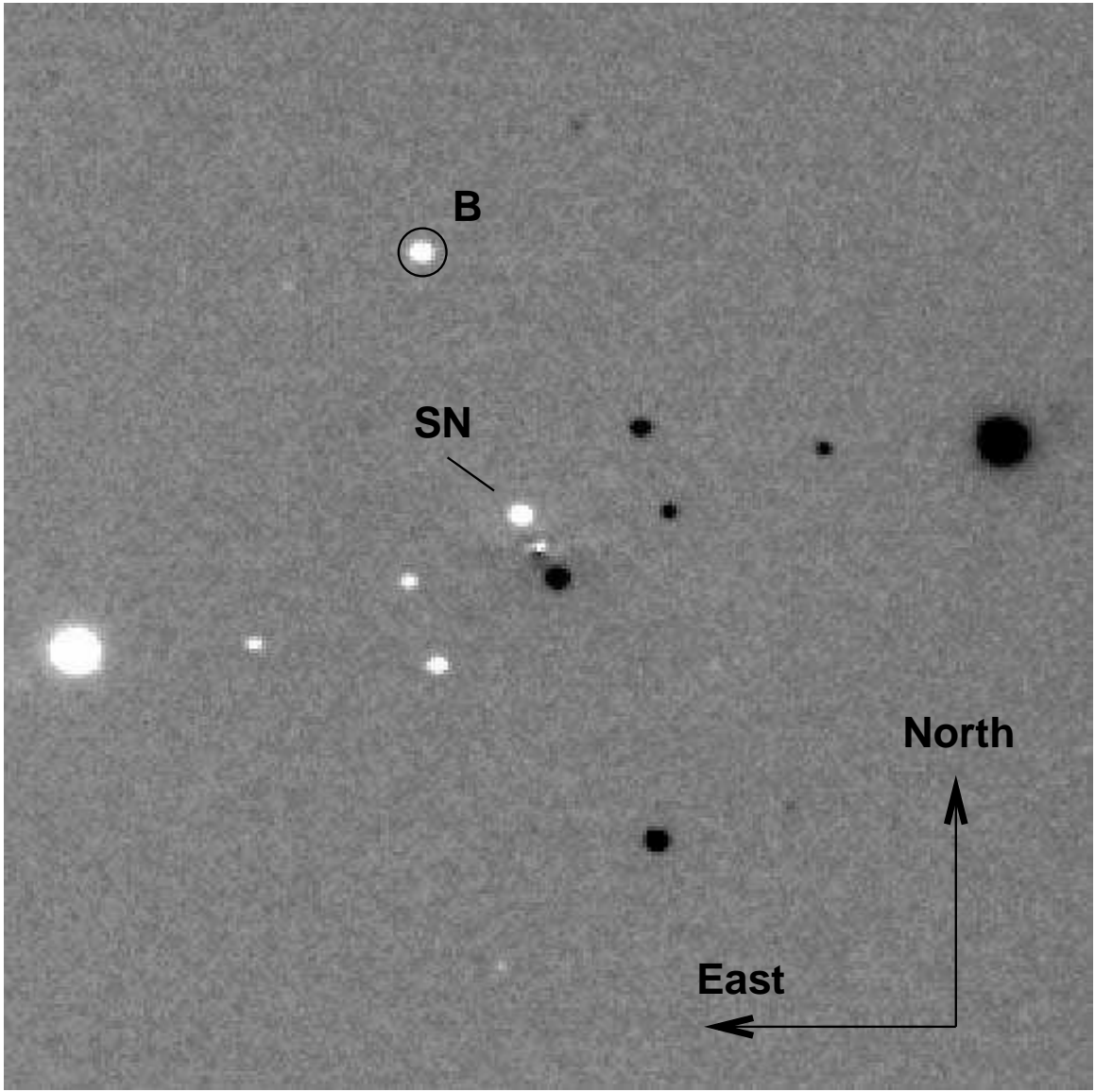}
\caption{The residual after rotation and subtraction
         of an image from RIT taken UT 2016 June 24.
         Positive residuals are bright and negative dark.
         North is up, East to the left;
         the galaxy's nucleus appears at the center.
         The field of view is roughly 7 by 7 arcminutes.
         \label{fig:rotsubrit} }
\end{figure}

The basic idea of the method
is to use the symmetry of the elliptical galaxy host
to provide a pseudo-template.
We identified the center of the host galaxy,
rotated the image by $180^\circ$ around this point,
then subtracted the rotated version from 
the original.
An example of the results, 
starting with the same image shown in 
Figure \ref{fig:chartritlabel},
is displayed in
Figure \ref{fig:rotsubrit};
we have zoomed in to show details near the nucleus more clearly.
The subtraction is not perfect:
small positive and negative residuals remain near
the center of the galaxy.
However, the residuals decrease rapidly with radius,
and near the position of the supernova
they are typically much smaller than the peak
of the supernova's light.
Moreover, the background
around and underneath the supernova is much
more uniform than in the original image,
removing the main source of error in the 
aperture photometry.

After creating these residual images,
we performed standard aperture photometry
of the SN and reference stars,
using the XVista 
\citep{Tref1989}
routines 
{\tt stars} 
and
{\tt phot}.
We chose to measure light
within
circular apertures of a fixed radius,
a bit larger than the usual FWHM:
4 pixels = 
$5{\rlap.}^{''}5$.
A local sky background was estimated for each star
using an annulus with radii of 
$6{\rlap.}^{''}9$
and
$13{\rlap.}^{''}8$.

\begin{figure}
\plotone{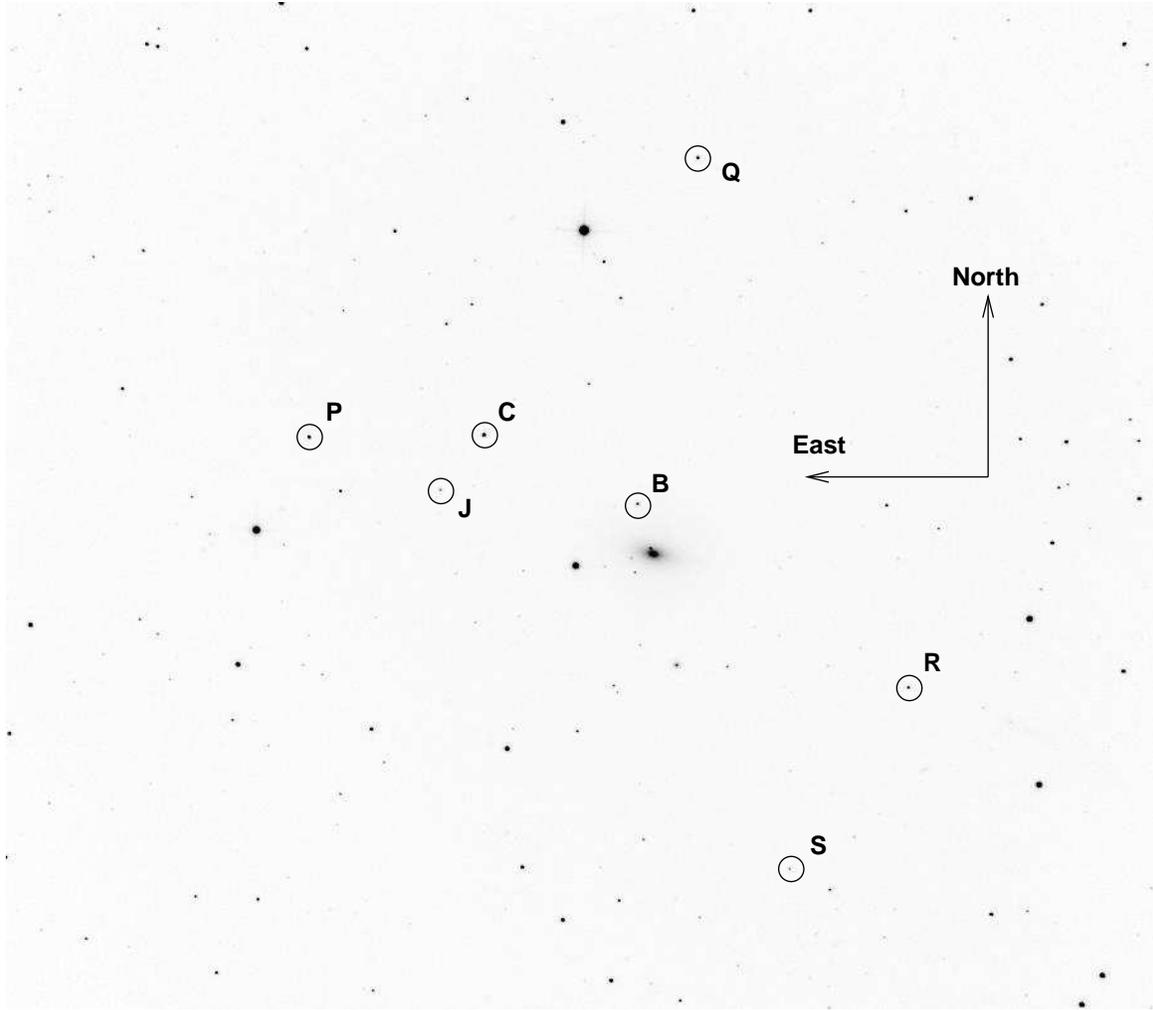}
\caption{A R-band composite (5 images of 45 seconds each)
         of SN 2016coj from NSO,
         showing stars used to calibrate measurements.
         North is up, East to the left.
         The field of view is roughly 30 by 30 arcminutes.
         \label{fig:chartvietjelabel} }
\end{figure}

% Now the Vietje details
The Northern Skies Observatory is located in Peacham, Vermont,
at longitude 72:09:57 West,
latitude +44:19:30 North,
and an elevation of 384 meters above sea level.
Images of SN 2016coj were acquired through a
43-cm f/6.8 corrected Dall-Kirkham astrograph made
by PlaneWave Instruments.
Light passes through Johnson-Cousins BVRI filters before
reaching an Apogee Alta U16M CCD camera;
we bin the chip 2x2 to produce a plate scale
of 
$1{\rlap.}^{''}26$
per pixel.
We acquire new flatfield images for each observing session,
but re-use bias and dark frames for a month or so.
We acquired 5 unguided images in each passband,
using exposure times of 45 to 60 seconds each.
After using MaximDL to subtract master bias and master dark frames,
and divide by a master flatfield frame,
we combined the images in each passband using a median technique.
These combined images typically had a FWHM 
ranging between
$3{\rlap.}^{''}1$
and
$3{\rlap.}^{''}9$,
with most lying near the low end of this range.
A sample composite $R$-band image is shown
in
Figure \ref{fig:chartvietjelabel}.

\begin{figure}
\plotone{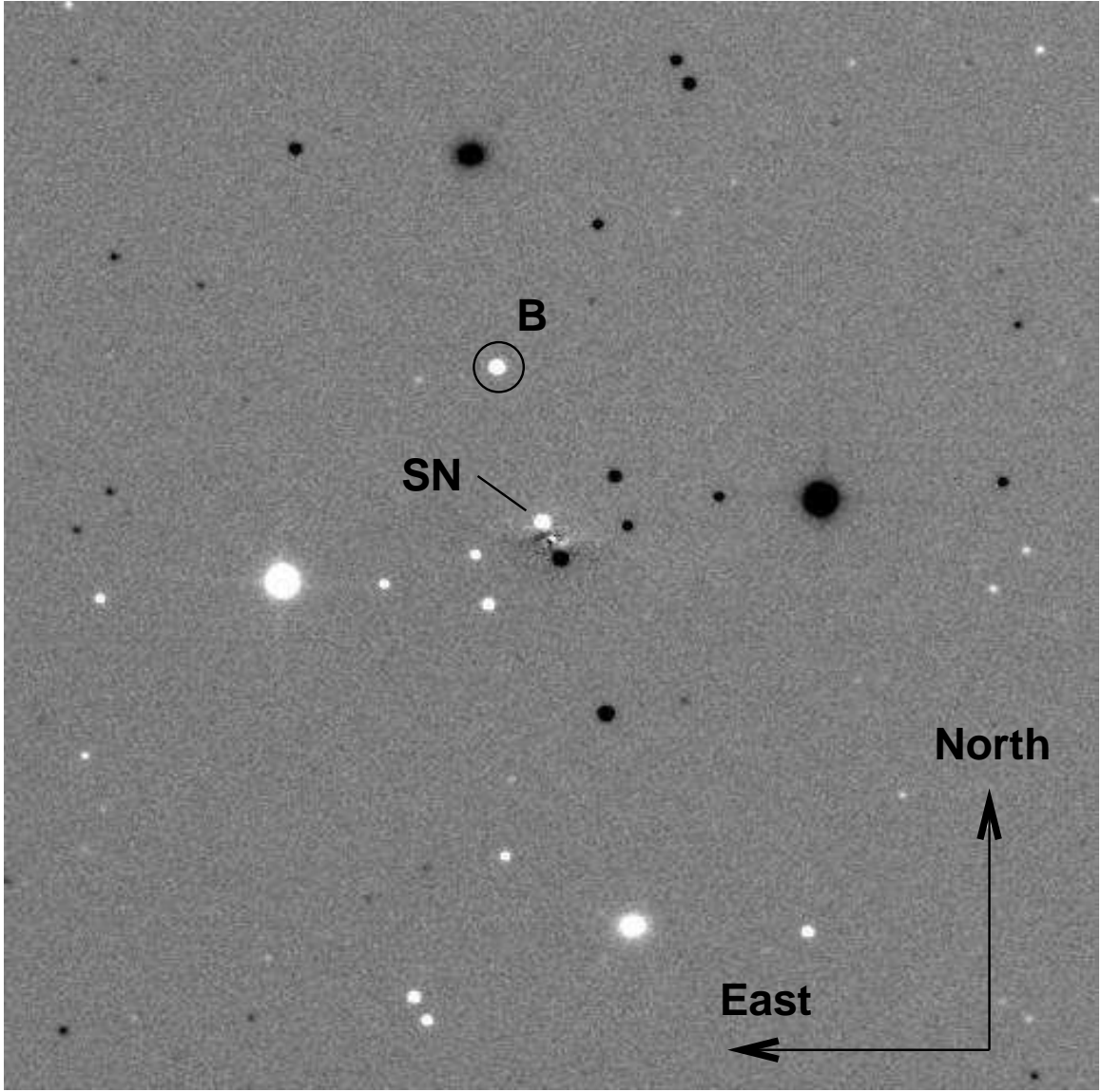}
\caption{The residual after rotation and subtraction
         of an image from NSO taken UT 2016 June 26.
         Positive residuals are bright and negative dark.
         North is up, East to the left;
         the galaxy's nucleus appears at the center.
         The field of view is roughly 10 by 10 arcminutes.
         \label{fig:rotsubvietje} }
\end{figure}

We applied the rotation technique described above 
to each combined image before extracting 
photometry using circular apertures of
radius
3 pixels =
$3{\rlap.}^{''}8$.
The local background was measured for each star
using an annulus of radii
$12{\rlap.}^{''}6$
and
$25{\rlap.}^{''}2$.
Figure \ref{fig:rotsubvietje}
shows one such residual image 
from the NSO dataset.

\section{Photometric calibration}

In order to transform our instrumental measurements
onto the standard Johnson-Cousins BVRI system,
we used a set of local reference stars provided
by the AAVSO\footnote
{American Association of Variable Star Observers,
  {\url{http://www.aavso.org}} }
in their sequence X18345FX.
These stars are labelled in all figures showing
the supernova and its surroundings.
Given our relatively small fields of view and 
shallow limiting magnitudes, we did not select comparison stars
on the basis of color, but accepted them all.
The color range covered is relatively small:
$0.598 \leq (B-V) \leq 1.065.$
SN 2016coj has a color of $(B-V) \simeq 0.0$
near maximum light, and does not redden
to match the comparison stars until about 
15 days later.
Of course, the spectrum of this type Ia SN is 
so distinct from that of the comparison stars
that color corrections must be approximate
in any case.

\begin{center}
 \begin{table*}[ht]
   \caption{Photometry of comparison stars}
   \label{tab:compstars}
   {\small
    \hfill{}
    \begin{tabular}{l l l c c c c}
    \hline
    Star & RA (J2000) &  Dec (J2000) &  B  & V & R & I  \\
    \hline

    B  &  12:08:11.72 & +65:12:04.7 & $15.198 \pm 0.086$ & $14.133 \pm 0.052$ & $13.627 \pm 0.116$ & $13.155 \pm 0.156$ \\
    C  &  12:09:01.20 & +65:14:03.5 & $13.317 \pm 0.093$ & $12.673 \pm 0.058$ & $12.316 \pm 0.121$ & $11.980 \pm 0.161$ \\
    J  &  12:09:13.84 & +65:12:09.7 & $15.607 \pm 0.109$ & $14.956 \pm 0.065$ & $14.603 \pm 0.136$ & $14.271 \pm 0.182$ \\
    K  &  12:09:00.39 & +65:07:51.9 & $16.573 \pm 0.123$ & $15.975 \pm 0.082$ & $15.547 \pm 0.174$ & $15.147 \pm 0.231$ \\
    P  &  12:09:56.10 & +65:13:37.8 & $13.778 \pm 0.095$ & $13.113 \pm 0.058$ & $12.750 \pm 0.116$ & $12.409 \pm 0.154$ \\
    Q  &  12:08:11.72 & +65:12:04.7 & $15.198 \pm 0.086$ & $14.133 \pm 0.052$ & $13.627 \pm 0.116$ & $13.155 \pm 0.156$ \\
    R  &  12:06:43.72 & +65:06:30.3 & $15.392 \pm 0.097$ & $14.424 \pm 0.058$ & $13.911 \pm 0.119$ & $13.432 \pm 0.158$ \\
    S  &  12:07:17.93 & +65:00:19.8 & $15.929 \pm 0.108$ & $15.246 \pm 0.065$ & $14.869 \pm 0.128$ & $14.515 \pm 0.169$ \\

    \hline
    \hline
    % \multicolumn{7}{l}{\footnotesize $^{a}$ J2000 } \\
    % \hline
  \end{tabular}
  }
 \end{table*}
\end{center}

In order to convert the RIT measurements to the Johnson-Cousins system,
we analyzed images of the standard fields
PG1633+009 
and 
PG2213-006
\citep{Land1992}
taken on five nights between June and August, 2016.
Comparing our instrumental values to the standard magnitudes,
we determined transformation equations 
\begin{eqnarray}
  B  &=  b  +  0.2016 (0134) * (b - v)  +  Z_B   \\
  V  &=  v  -  0.0920 (0063) * (v - r)  +  Z_V   \\
  R  &=  r  -  0.1137 (0058) * (r - i)  +  Z_R   \\
  I  &=  i  -  0.0174 (0034) * (r - i)  +  Z_I   
\end{eqnarray}
In the equations above, 
lower-case symbols represent instrumental magnitudes,
upper-case symbols Johnson-Cousins magnitudes,
terms in parentheses the uncertainties in each coefficient,
and $Z$ the zeropoint in each band.
The relatively small field of view of 
the RIT images allowed us to use 
only a few stars for calibration:
B, C, J, and, in some cases, K.

In mid-June 2016,
we noticed that images in the $B$-band from
RIT had a low signal-to-noise ratio,
even after coaddition;
this is largely a function of the relatively low
sensitivity of our camera's sensor at short wavelengths.
The $B$-band measurements showed a large
scatter from night to night,
making real trends in the light curve hard to discern.
Therefore, 
after UT 2016 June 20,
we stopped taking images at RIT in the $B$-band.

We present
our calibrated measurements of SN 2016coj made at RIT
in 
Table \ref{tab:ritphot}.
The first column shows the mean Julian Date of all the exposures
taken during each night;
we have subtracted the arbitrary constant
2457530 from all Julian Dates for convenience.
%  Contact the first author for a dataset providing the Julian Date 
%  of each measurement individually.
The uncertainties listed in Table
\ref{tab:ritphot}
incorporate the uncertainties in instrumental magnitudes
and in the offset to shift the instrumental values
to the standard scale, added in quadrature.

\begin{center}
 \begin{deluxetable}{l l l l l l}
   \tablecaption{RIT photometry of SN 2016coj \label{tab:ritphot}}
   \tablehead{
     \colhead{JD-2457530} &
     \colhead{B} &
     \colhead{V} &
     \colhead{R} &
     \colhead{I} &
     \colhead{comments}
   }

\startdata
     9.64  &  $14.108 \pm 0.040$  &  $13.960 \pm 0.029$  &  $13.811 \pm 0.067$  &  $13.866 \pm 0.085$  &  \\  
    10.61  &  $13.816 \pm 0.042$  &  $13.760 \pm 0.021$  &  $13.657 \pm 0.020$  &  $13.677 \pm 0.040$  &  \\  
    11.61  &  $13.663 \pm 0.067$  &  $13.556 \pm 0.028$  &  $13.432 \pm 0.104$  &  $13.484 \pm 0.125$  &  cirrus \\  
    19.60  &  $13.231 \pm 0.062$  &  $13.072 \pm 0.038$  &  $13.061 \pm 0.021$  &  $13.426 \pm 0.073$  &  \\  
    20.59  &  $13.371 \pm 0.090$  &  $13.109 \pm 0.057$  &  $13.070 \pm 0.038$  &  $13.503 \pm 0.105$  &  cirrus \\  
    21.58  &  $13.244 \pm 0.077$  &  $13.139 \pm 0.049$  &  $13.161 \pm 0.040$  &  $13.548 \pm 0.049$  &  light clouds \\  
    22.64  &  $13.368 \pm 0.058$  &  $13.196 \pm 0.057$  &  $13.268 \pm 0.077$  &  $13.822 \pm 0.136$  &  \\  
    23.59  &  $13.415 \pm 0.027$  &  $13.179 \pm 0.029$  &  $13.212 \pm 0.021$  &  $13.714 \pm 0.059$  &  \\  
    24.59  &  $13.480 \pm 0.049$  &  $13.232 \pm 0.029$  &  $13.369 \pm 0.046$  &  $13.837 \pm 0.101$  &  cirrus \\  
    27.60  &  $13.884 \pm 0.045$  &  $13.409 \pm 0.027$  &  $13.544 \pm 0.061$  &  $13.970 \pm 0.065$  &  \\  
    28.60  &  $13.998 \pm 0.050$  &  $13.440 \pm 0.022$  &  $13.632 \pm 0.059$  &  $13.984 \pm 0.063$  &  cirrus \\  
    29.61  &  $13.975 \pm 0.096$  &  $13.472 \pm 0.030$  &  $13.543 \pm 0.065$  &  $13.859 \pm 0.088$  &  bright moon \\  
    31.61  &  $14.270 \pm 0.072$  &  $13.586 \pm 0.028$  &  $13.648 \pm 0.079$  &  $13.719 \pm 0.111$  &  \\  
    32.59  &  $14.436 \pm 0.049$  &  $13.639 \pm 0.023$  &  $13.648 \pm 0.072$  &  $13.750 \pm 0.085$  &  \\  
    33.60  &  $14.535 \pm 0.088$  &  $13.650 \pm 0.024$  &  $13.675 \pm 0.057$  &  $13.723 \pm 0.087$  &  \\  
    36.59  &  $15.166 \pm 0.222$  &  $13.865 \pm 0.048$  &  $13.656 \pm 0.088$  &  $13.608 \pm 0.085$  &  clouds \\  
    37.59  &  $15.129 \pm 0.091$  &  $13.964 \pm 0.052$  &  $13.670 \pm 0.066$  &  $13.572 \pm 0.099$  &  fewer images \\  
    39.60  &  $15.400 \pm 0.115$  &  $14.061 \pm 0.030$  &  $13.737 \pm 0.069$  &  $13.516 \pm 0.086$  &  \\  
    40.60  &  $15.370 \pm 0.069$  &  $14.155 \pm 0.043$  &  $13.795 \pm 0.061$  &  $13.418 \pm 0.104$  &  \\  
    42.60  &  $15.678 \pm 0.119$  &  $14.349 \pm 0.044$  &  $13.948 \pm 0.072$  &  $13.593 \pm 0.086$  &  \\  
    43.60  &  $15.722 \pm 0.078$  &  $14.421 \pm 0.040$  &  $14.034 \pm 0.081$  &  $13.705 \pm 0.095$  &  \\  
    45.60  &  $15.938 \pm 0.079$  &  $14.586 \pm 0.044$  &  $14.248 \pm 0.078$  &  $13.840 \pm 0.082$  &  haze \\  
    50.60  &  $16.026 \pm 0.117$  &  $14.822 \pm 0.069$  &  $14.462 \pm 0.093$  &  $14.372 \pm 0.106$  &  cirrus \\  
    51.60  &  $15.844 \pm 0.189$  &  $14.897 \pm 0.049$  &  $14.578 \pm 0.085$  &  $14.272 \pm 0.113$  &  cirrus \\  
    53.60  &  $16.265 \pm 0.107$  &  $14.926 \pm 0.062$  &  $14.654 \pm 0.081$  &  $14.414 \pm 0.107$  &  haze, old flats \\  
    56.60  &  $16.193 \pm 0.117$  &  $15.100 \pm 0.055$  &  $14.799 \pm 0.082$  &  $14.596 \pm 0.098$  &  \\  
    58.60  &  $16.002 \pm 0.113$  &  $15.111 \pm 0.058$  &  $14.796 \pm 0.086$  &  $14.609 \pm 0.113$  &  \\  
    59.60  &  $16.202 \pm 0.102$  &  $15.069 \pm 0.047$  &  $14.815 \pm 0.064$  &  $14.623 \pm 0.084$  &  start longer exp \\  
    62.60  &  \nodata \nodata  &  $15.188 \pm 0.035$  &  $14.915 \pm 0.084$  &  $14.686 \pm 0.110$  &  \\  
	 63.61  &  \nodata \nodata  &  $15.222 \pm 0.063$  &  $15.032 \pm 0.090$  &  $14.824 \pm 0.124$  &  cirrus \\  
    65.60  &  \nodata \nodata  &  $15.257 \pm 0.038$  &  $15.020 \pm 0.062$  &  $14.840 \pm 0.082$  &  \\  
    66.60  &  \nodata \nodata  &  $15.289 \pm 0.049$  &  $15.061 \pm 0.082$  &  $14.839 \pm 0.125$  &  cirrus \\  
    69.62  &  \nodata \nodata  &  $15.374 \pm 0.041$  &  $15.176 \pm 0.087$  &  $14.928 \pm 0.083$  &  light clouds \\  
    72.60  &  \nodata \nodata  &  $15.433 \pm 0.036$  &  $15.211 \pm 0.083$  &  $14.949 \pm 0.089$  &  \\  
    73.59  &  \nodata \nodata  &  $15.451 \pm 0.039$  &  $15.235 \pm 0.069$  &  $15.118 \pm 0.090$  &  \\  
    75.61  &  \nodata \nodata  &  $15.503 \pm 0.092$  &  $15.421 \pm 0.108$  &  $15.054 \pm 0.126$  &  light clouds \\  
    \enddata
 \end{deluxetable}
\end{center}

We determined linear transformations between
the instrumental NSO measurements
and the standard scale using images of
the open cluster M67
and photometry provided by the AAVSO.
The transformation equations for NSO were
\begin{eqnarray}
  B  &=  b  -  0.164 (0.033) * (b - v)  +  Z_B   \\
  V  &=  v  -  0.109 (0.023) * (b - v)  +  Z_V   \\
  V  &=  v  -  0.197 (0.050) * (v - r)  +  Z_V   \\
  R  &=  r  -  0.205 (0.052) * (r - i)  +  Z_R   \\
  I  &=  i  -  0.238 (0.073) * (r - i)  +  Z_I   
\end{eqnarray}
In the equations above, 
lower-case symbols represent instrumental magnitudes,
upper-case symbols Johnson-Cousins magnitudes,
terms in parentheses the uncertainties in each coefficient,
and $Z$ the zeropoint in each band.
We list two equations for the $V$-band;
on nights when we acquired $R$ images,
we used the $(v - r)$ equation;
but on nights when we measured only $B$ and $V$,
we used the $(b - v)$ version.

Table \ref{tab:nsophot} lists our calibrated measurements 
of SN 2016coj made at Northern Skies Observatory.

\begin{center}
 \begin{deluxetable}{l l l l l l}
   \tablecaption{NSO photometry of SN 2016coj \label{tab:nsophot} }
   \tablehead{
     \colhead{JD-2457530} &
     \colhead{B} &
     \colhead{V} &
     \colhead{R} &
     \colhead{I} &
     \colhead{comments}
   }

\startdata
    20.64  &  $13.307 \pm 0.023$  &  $13.073 \pm 0.027$  &  $13.121 \pm 0.037$  &  $13.570 \pm 0.057$  &  \\  
    25.64  &  $13.606 \pm 0.056$  &  $13.242 \pm 0.015$  &  \nodata \nodata  &  \nodata \nodata  &  \\  
    33.65  &  $14.565 \pm 0.038$  &  $13.791 \pm 0.017$  &  $13.773 \pm 0.035$  &  $13.967 \pm 0.056$  &  \\  
    34.63  &  $14.682 \pm 0.025$  &  $13.845 \pm 0.026$  &  $13.769 \pm 0.034$  &  $13.824 \pm 0.047$  &  \\  
    35.64  &  $14.785 \pm 0.058$  &  $13.807 \pm 0.032$  &  \nodata \nodata  &  \nodata \nodata  &  \\  
    39.61  &  $15.195 \pm 0.067$  &  $14.037 \pm 0.032$  &  \nodata \nodata  &  \nodata \nodata  &  \\  
    42.61  &  $15.487 \pm 0.034$  &  $14.369 \pm 0.028$  &  $14.000 \pm 0.036$  &  $13.687 \pm 0.065$  &  \\  
    43.66  &  $15.626 \pm 0.047$  &  $14.461 \pm 0.030$  &  $14.110 \pm 0.039$  &  $13.729 \pm 0.058$  &  \\  
    45.59  &  $15.729 \pm 0.043$  &  $14.632 \pm 0.031$  &  \nodata \nodata  &  \nodata \nodata  &  \\  
    73.59  &  $16.423 \pm 0.058$  &  $15.590 \pm 0.038$  &  $15.456 \pm 0.042$  &  $15.535 \pm 0.075$  &  \\  
    \enddata
 \end{deluxetable}
\end{center}

\section{Light curves}

%  The very early observations
%  and detailed modeling reported in
%  \citet{Zhen2016}
%  yield a date of ``first light'' of
%  JD 2457531.33;
%  we will adopt this date as a proxy 
%  for the time of explosion in discussions below.
%  Note that this corresponds to a time of
%  1.33 days with our convention of subtracting
%  the large constant 2457530 from all Julian Dates
%  in our graphs.

\begin{figure}
 \plotone{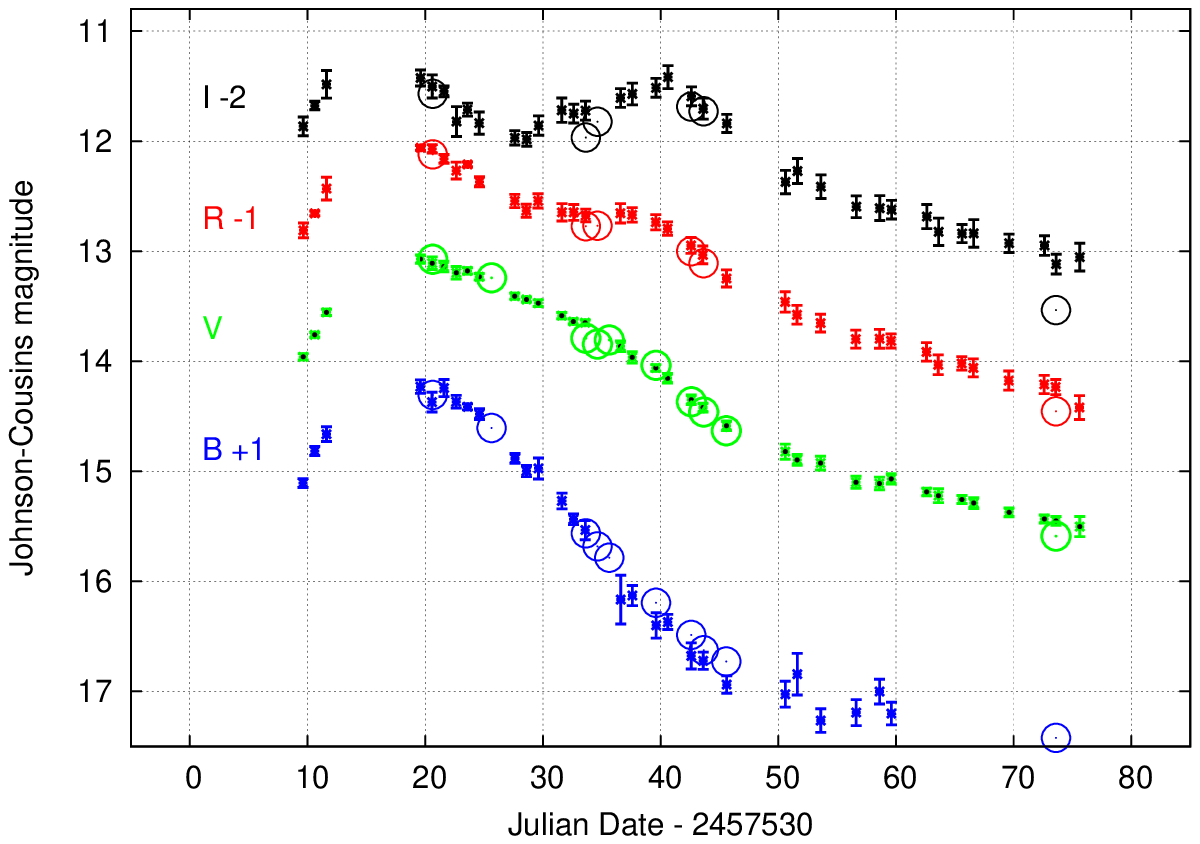}
 \caption{Light curves of SN 2016coj in BVRI.
          The data for each passband have
          been offset vertically for clarity.
          Small symbols represent measurements from RIT,
          large symbols those from NSO;
          uncertainties in the latter are smaller than the symbols.
          \label{fig:bvricurves} }
\end{figure}

In order to determine the time and magnitude at 
peak brightness,
we fit polynomials of order 3 
to the light curves near maximum,
using data from the period
$0 < (JD - 2457530) < 30$
in each passband,
weighting the fits by the uncertainties in each measurement.
We list the results in 
Table \ref{tab:appmax}.
For the secondary maximum in $I$-band,
we found that polynomials of order 2 provided
better fits;
we averaged the results from several intervals during
during the period
$30 < (JD - 2457530) < 50$
to produce the value in the table.
Note that the $I$-band magnitude
at its primary maximum is particularly uncertain,
as it falls farther within the gap in our measurements
than the peaks in other passbands.

Using a second-order polynomial to interpolate
in the B-band observations exactly 15 days
after the time of B-band maximum light,
we compute 
${\Delta}_{15}(B) = 1.32 \pm 0.10$.
By this measure, 
SN 2016coj lies in the range of ``normal''
type Ia SNe, 
such as 
1980N
\citep{Hamu1991},
1989B
\citep{Well1994},
1994D
\citep{Rich1995},
2003du
\citep{Stan2007},
and 2011fe
(\citealt{Rich2012}; \citealt{Parr2012}).
The 
secondary
peak in $I$-band,
which lies 
$22.8 \pm 1.0$ days after
and $0.27 \pm 0.07$ mag below the primary peak,
is also typical of ``normal'' type Ia events.

Our values of the apparent magnitude at $B$-band
maximum light 
and the
${\Delta}_{15}(B)$ parameter
agree with those measured by
\citet{Zhen2016}.
Those authors also provide spectroscopic
evidence
to support a 
``normal'' classification for SN 2016coj.

\begin{center}
 \begin{deluxetable}{l c r }
   \tablecaption{Apparent magnitudes at maximum light \label{tab:appmax} }
   \tablehead{
     \colhead{Passband} &
     \colhead{JD-2457530} &
     \colhead{mag}
   }

   \startdata
    B        &  $18.1  \pm 0.4$  &   $13.16 \pm 0.07$ \\
    V        &  $18.7  \pm 0.2$  &   $13.06 \pm 0.01$ \\
    R        &  $18.0  \pm 0.4$  &   $13.04 \pm 0.03$ \\
    I        &  $16.2  \pm 2.9$  &   $13.23 \pm 0.10$ \\
    I (sec)  &  $39.0  \pm 0.7$  &   $13.50 \pm 0.05$ \\
    \enddata
 \end{deluxetable}
\end{center}

% need to derive extinction here, so I can correct colors for reddening
In order to compute absolute magnitudes and intrinsic colors
for SN 2016coj,
we must remove the extinction due to any intervening material.
Fortunately, there appears to be very little dust
in its direction.
Our own Galaxy's contribution is small:
\citet{Schl1998}
use infrared maps of the Milky Way to estimate
$E(B-V)_{\rm MW} = 0.017$
in the direction of NGC 4125.
\citet{Zhen2016} examine high-resolution spectra
of SN 2016coj to look for absorption lines
caused by interstellar material in the host galaxy.
Finding none, they employ several methods to place 
upper limits on the reddening of 
$E(B-V)_{\rm host} \lesssim 0.05$ or $E(B-V)_{\rm host} \lesssim 0.09$.
We will adopt a host value of
$E(B-V)_{\rm host} = 0.05$ for the color curves 
we present below,
yielding a total reddening of
$E(B-V)_{\rm tot} = 0.067$.
Following the conversions from reddening to extinction
given in
\citet{Schl2011},
we derive the extinction to SN 2016coj to be
$A_B = 0.24$,
$A_V = 0.18$,
$A_R = 0.15$,
and
$A_I = 0.10$.

\begin{figure}
 \plotone{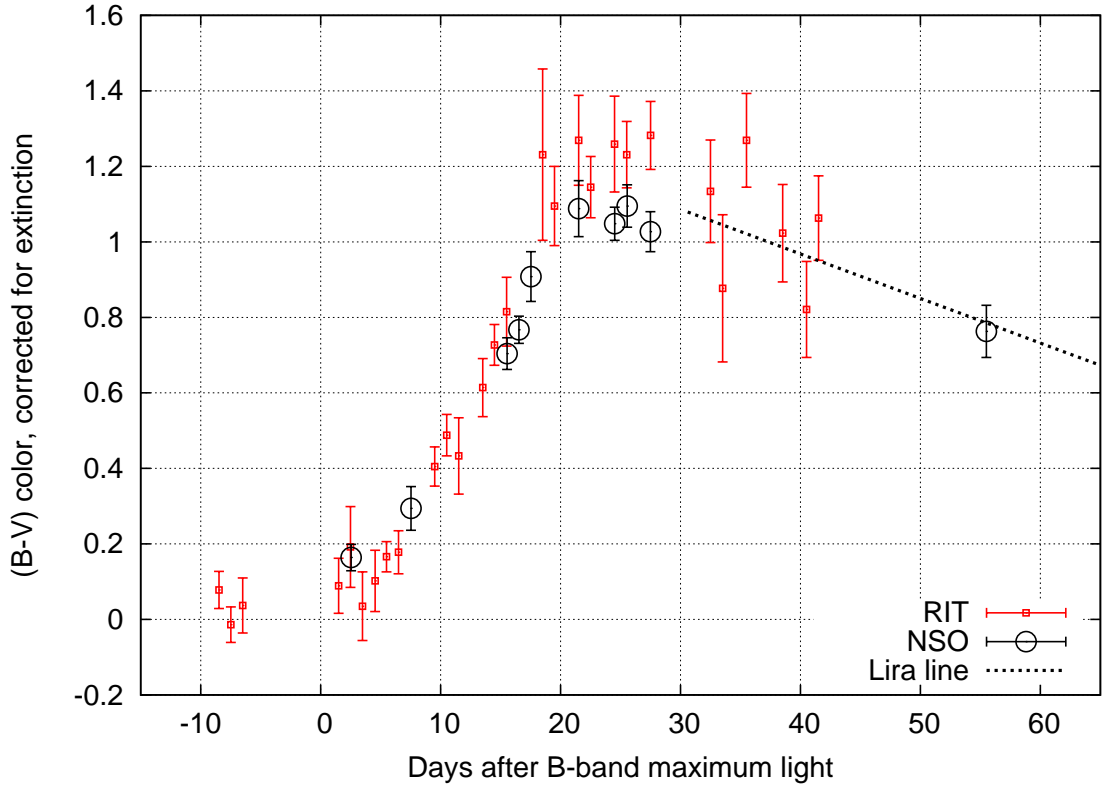}
 \caption{(B-V) color evolution of SN 2016coj,
          after correcting for extinction.
          \label{fig:bvcolor} }
\end{figure}

\begin{figure}
 \plotone{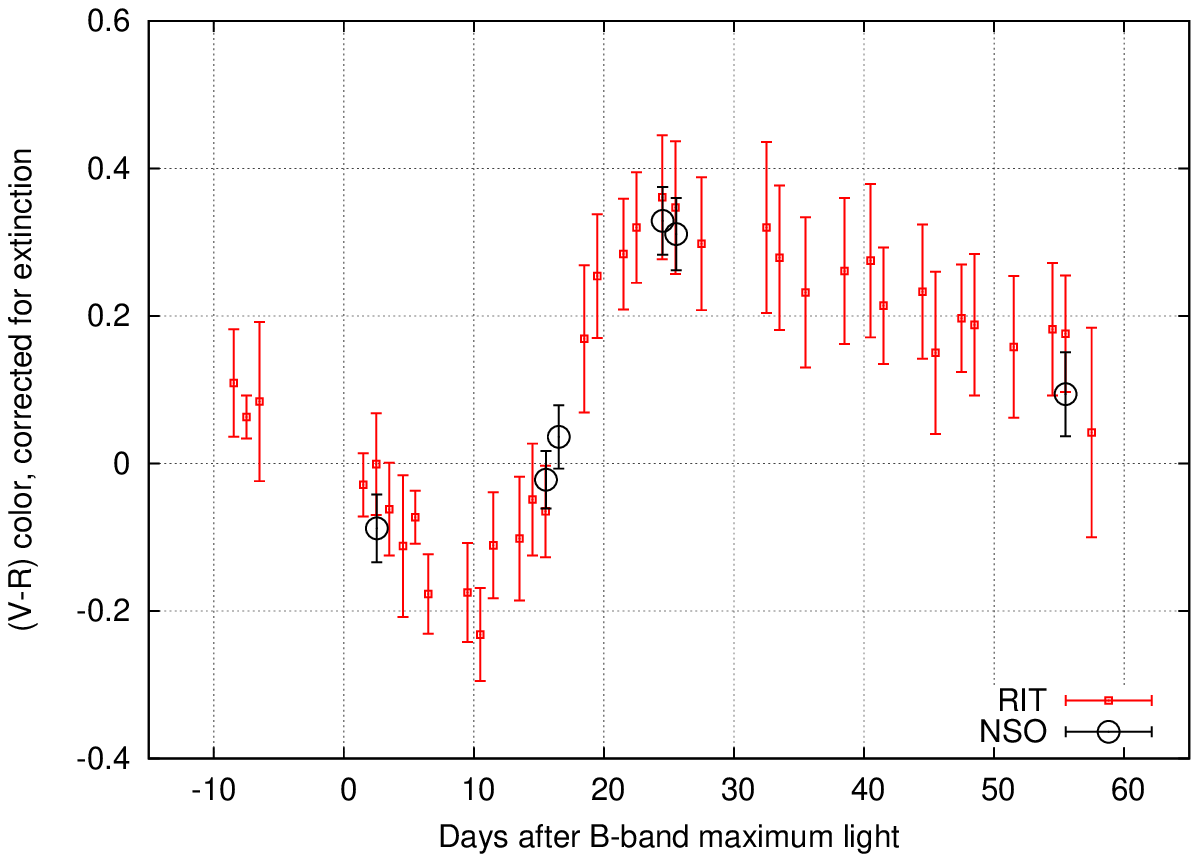}
 \caption{(V-R) color evolution of SN 2016coj,
          after correcting for extinction.
          \label{fig:vrcolor} }
\end{figure}

\begin{figure}
 \plotone{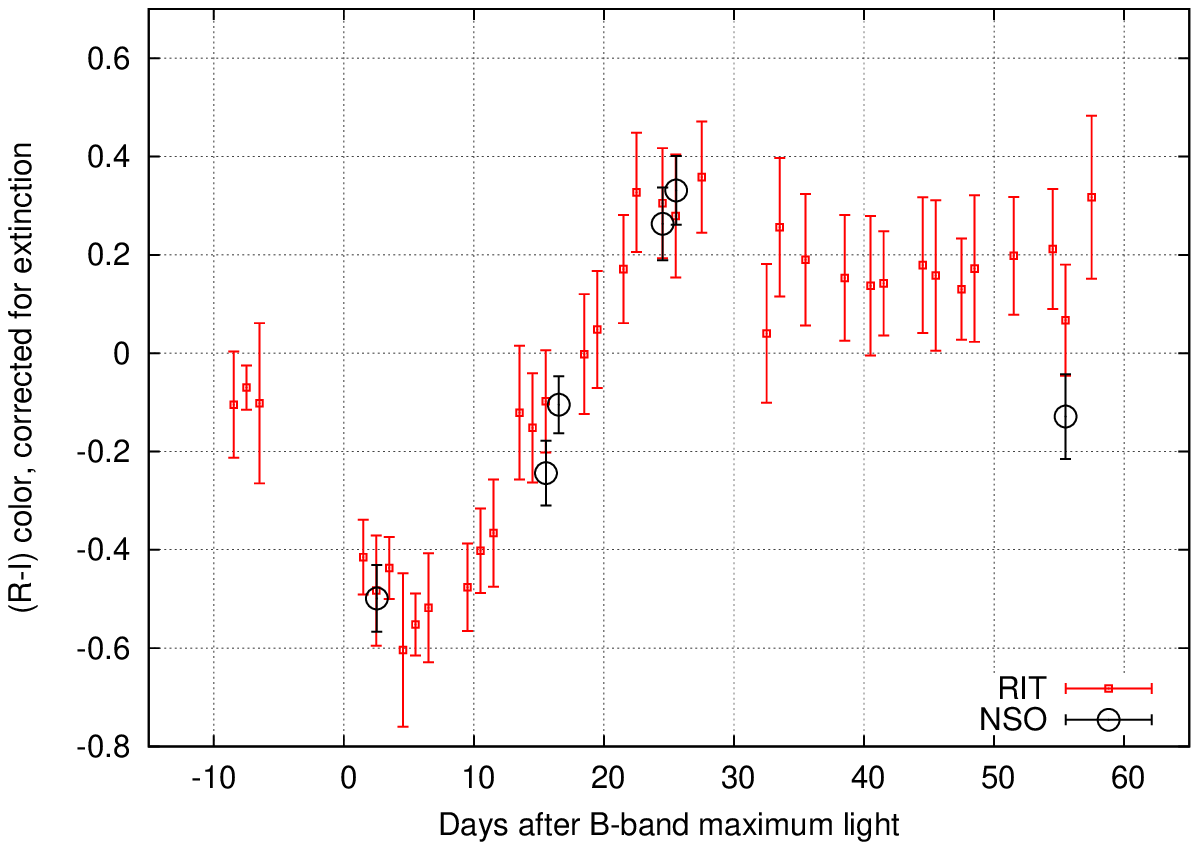}
 \caption{(R-I) color evolution of SN 2016coj,
          after correcting for extinction.
          \label{fig:ricolor} }
\end{figure}

After removing this extinction from each passband,
we calculate the evolution of the event in each color;
see 
Figure 
\ref{fig:bvcolor} for $(B-V)$,
Figure
\ref{fig:vrcolor} for $(V-R)$,
and
Figure
\ref{fig:ricolor} for $(R-I)$.
The $(B-V)$ color shows a value of zero
at maximum light,
typical for a normal type Ia.
In the same figure, we have drawn a line 
which represents the late-time $(B-V)$ evolution of 
a set of normal type Ia SNe with little or no extinction
(\citet{Lira1995} ; \citet{Phil1999}).
Although our measurements are sparse and noisy
at late times, due to the low signal in the $B$ band,
they suggest that SN 2016coj followed the same
evolution as other normal events.
In Figure
\ref{fig:vrcolor},
we see that SN 2016coj reaches a minimum $(V-R)$ color 
about 10 days after $B$-band maximum, 
then increases to a maximum $(V-R) = 0.35$.
The time of minimum is a few days earlier in $(R-I)$,
which then rises to a maximum of $(R-I) = 0.35.$
All these properties are similar to those in the color curves
of the normal SNe Ia
1994D
\citep{Rich1995},
2003du
\citep{Stan2007},
2009an
\citep{Sahu2013},
and
2011fe
\citep{Rich2012}.
The only significant difference in the 
late-time
behavior of SN 2016coj is in $(R-I)$,
which appears to have a relatively constant 
value of $(R-I) \sim 0.2$.
However, we believe that this color in particular
suffers from a systematic bias in the RIT
$I$-band measurements (see the Appendix);
note the position of the single late-time NSO
datum, at a negative color more typical of
normal SNe.

\section{Absolute magnitudes}

What was the absolute magnitude of SN 2016coj?
In order to convert our apparent magnitudes to the absolute scale,
we need to account for extinction and the distance to the host galaxy.
As mentioned earlier, 
\citet{Zhen2016}
place only upper limits on the extinction due to material
in the host galaxy.
In the discussion which follows, we will compute two values 
of absolute magnitude, one assuming no extinction in the host galaxy,
the other corresponding to the upper limit of $E(B-V) = 0.05$
derived in 
\citet{Zhen2016}.
The distance to NGC 4125 has been measured a number of times,
but none are very recent.  
We will adopt the measurement based on Surface Brightness Fluctuations (SBF)
by
\citet{Tonr2001}
of 
$(m - M) = 31.89 \pm 0.25.$

\begin{center}
 \begin{deluxetable}{l r r r}
   \tablecaption{Absolute magnitudes of SN 2016coj at maximum light, corrected for extinction \label{tab:absmax} }
   \tablehead{
     \colhead{Passband} &
     \colhead{$E(B-V) = 0$ in host} &
     \colhead{$E(B-V) = 0.05$ in host } &
     \colhead{based on $\Delta m_{15}(B)$ \tablenotemark{a} } 
   }

   \startdata
    B         &   $-18.79 \pm 0.26$ \phantom{xxx}  &  $-18.97 \pm 0.26$  \phantom{xxx} & $-19.19 \pm 0.10$ \phantom{xxx} \\
    V         &   $-18.88 \pm 0.26$ \phantom{xxx}  &  $-19.01 \pm 0.26$  \phantom{xxx} & $-19.12 \pm 0.09$ \phantom{xxx} \\
    R         &   $-18.89 \pm 0.26$ \phantom{xxx}  &  $-18.99 \pm 0.26$  \phantom{xxx} & $-19.14 \pm 0.07$ \phantom{xxx} \\
    I         &   $-18.68 \pm 0.27$ \phantom{xxx}  &  $-18.76 \pm 0.27$  \phantom{xxx} & $-18.87 \pm 0.08$ \phantom{xxx} \\
    I (sec)   &   $-18.41 \pm 0.26$ \phantom{xxx}  &  $-18.49 \pm 0.26$  \phantom{xxx} & \nodata  \phantom{xxxxxxx} \\
    \enddata
    \tablenotetext{a}{using the relationship from \citet{Prie2006} }
 \end{deluxetable}
\end{center}

A connection between the absolute magnitude
of a type Ia SN and its rate of decline after maximum
was first noted by
\citet{Phil1993}
and has since been refined by a number of authors
(\citealt{Hamu1996} ; \citealt{Ries1996} ; \citealt{Perl1997}).
We choose the relationships derived by 
\citet{Prie2006} 
which are based on the fading in $B$-band
in the first 15 days after maximum light,
the 
$\Delta m_{15}(B)$ parameter.
In the case of SN 2016coj,
we measure
$\Delta m_{15}(B) = 1.32 \pm 0.10$.
Inserting that into the equations in Table 3 of
\citet{Prie2006} 
for events in environments with low extinction,
we derive the absolute magnitudes 
shown in the rightmost column of 
Table \ref{tab:absmax}.
With the exception of the $B$-band measurement
assuming no host extinction,
all measurements agree with the predictions of the
decline-rate method,
supporting further the classification of SN 2016coj
as normal.
The slight improvement offered by assuming a 
small host extinction provides weak evidence
that it may be close to the upper limits
derived by
\citet{Zhen2016}.

\section{Conclusion}

Our measurements of SN 2016coj
show that its photometric behavior at early 
times (within 60 days of maximum light)
follows that of ``normal'' type Ia SNe.
We compute a decline parameter of
$\Delta m_{15}(B) = 1.32 \pm 0.10$ mag,
placing it in the middle of the distribution
of normal events.
Adopting a distance modulus to NGC 4125
of
$(m - M) = 31.89$
and correcting for a total of
$E(B-V) = 0.067$ of extinction,
we derive absolute magnitudes
of
$M_B = -19.01$,
$M_V = -19.05$,
$M_R = -19.03$
and 
$M_I = -18.79$.

We have shown that correcting for contamination of SN measurements
by the background light of the host galaxy is a 
difficult issue for this event.
While our measurements at early times -- upon which the 
above conclusions are based -- are reliable,
those at late times must be treated with caution.
The simple procedure we used for this dataset
does not require template images of the host galaxy alone,
but can leave a systematic error which grows as the
target object fades.

\acknowledgements
We thank the staff at AAVSO for 
their finding charts, sequences of comparison stars, 
and endless support of observers.
MWR is grateful for the continued support of the
RIT Observatory by RIT and its College of Science.
BV thanks the Northeast Kingdom Astronomy Foundation
for allowing him to use their facility.
The Lick Observatory Supernova Search noticed this 
event and quickly alerted the rest of the community,
permitting us to begin our study while the supernova
was still on the rise.
This research has made use of the ``Aladin Sky Atlas''
and SIMBAD database
\citep{Weng2000},
developed and operated at the CDS, Strasbourg Observatory, France,
We have also made use of the NASA/IPAC Extragalactic Database (NED)
which is operated by the Jet Propulsion Laboratory, 
California Institute of Technology, 
under contract with the National Aeronautics and Space Administration.
This paper (and many others) has made 
use of NASA's Astrophysics Data System.

\facility{AAVSO}

\appendix
\section{Bias in photometry}

The location of SN 2016coj in its host galaxy presents
the observer with good news and bad news.
The good news is that NGC 4125 is an elliptical galaxy,
and therefore likely contains relatively little gas and dust,
as the spectroscopic measurements of 
\citet{Zhen2016} 
confirm.
It is therefore not necessary to make large corrections
for extinction.
But the bad news is that the supernova is close enough
to the bright core of its host,
offset by only 
$5{\rlap.}^{''}0$ east
and 
$10{\rlap.}^{''}8$ north
\citep{Zhen2016},
that the light of the nucleus and the surrounding stars
provides a significant background 
to measurements of the SN.
Moreover, at this location,
the light of the galaxy has a strong radial gradient,
making it very difficult to subtract its contribution accurately.

Since other observers may encounter similar situations,
we describe in some detail below our investigation of 
the likely systematic errors that can arise
when one attempts to perform photometry on such images.

It became clear to us that simple aperture photometry methods
would yield poor results in this case.
Since we did not have template images of the galaxy to use
as references for subtraction in the standard manner,
we settled upon a method which would provide better
(but not perfect) results:
making a copy of each image, rotating the copy by 
$180^\circ$ around the center of the galaxy,
then subtracting the copy from the original.
As shown in 
Figure \ref{fig:rotsubrit},
the resulting residual image has a background near the
location of the supernova which is both much lower, and much more uniform,
than the original image.
We used these residual images to derive the measurements 
presented in this paper.

However, we suspected that there remained a sometimes significant
systematic error in the photometry produced by this method,
for two reasons.
First, when comparing the decline of SN 2016coj against that of
other type Ia SNe, such as SN 2011fe, we noticed that this event
faded less quickly at late times -- by an amount which appeared
to grow with time.
Second, we noticed a difference between measurements
from our two sites:
measurements from NSO, which typically have a smaller PSF 
and higher signal-to-noise than those from RIT, 
showed the SN as slightly fainter, and this difference also
grew with time.

Could there be a reason why measurements of faint point sources
immersed in a noisy (and possibly non-uniform) background
should show a systematic error?
Our technique of identifying and measuring both reference stars
and the supernova in images was a simple one:
we searched through each image independently to identify
peaks above the local sky background, measured their properties,
and kept as ``good'' sources those which had 
shapes consistent with the expected PSF.
We used the pixels around each of these sources to compute its
center,
placed a circular aperture at this position,
and integrated the light within the aperture;
finally, we subtracted the contribution from the local background
light within the aperture.

The important feature of this standard method is that
the position of the aperture used to measure 
the SN (and reference stars) is not fixed in any way:
it may be influenced independently by noise in each image,
especially when the source is faint and the noise is high.
A better technique, one which is natural when using a template,
is to align images, co-add them to improve signal-to-noise,
and measure relative positions for the SN and several reference stars;
then, in each individual image, measure the position of 
bright reference stars and use them to {\it infer} the
position of the SN using a fixed offset, rather than computing it based on
the possibly noisy data at its location in each image.
The method we employed is likely to shift the center
of the SN's aperture slightly to follow positive noise peaks,
which could yield measurements slightly higher than they ought to be.

In short, we suspect that our measurements,
especially those made at RIT, 
contain a systematic positive bias:
the SN appears brighter than it actually is,
by an amount which increases as the SN fades.

\begin{figure}
 \plotone{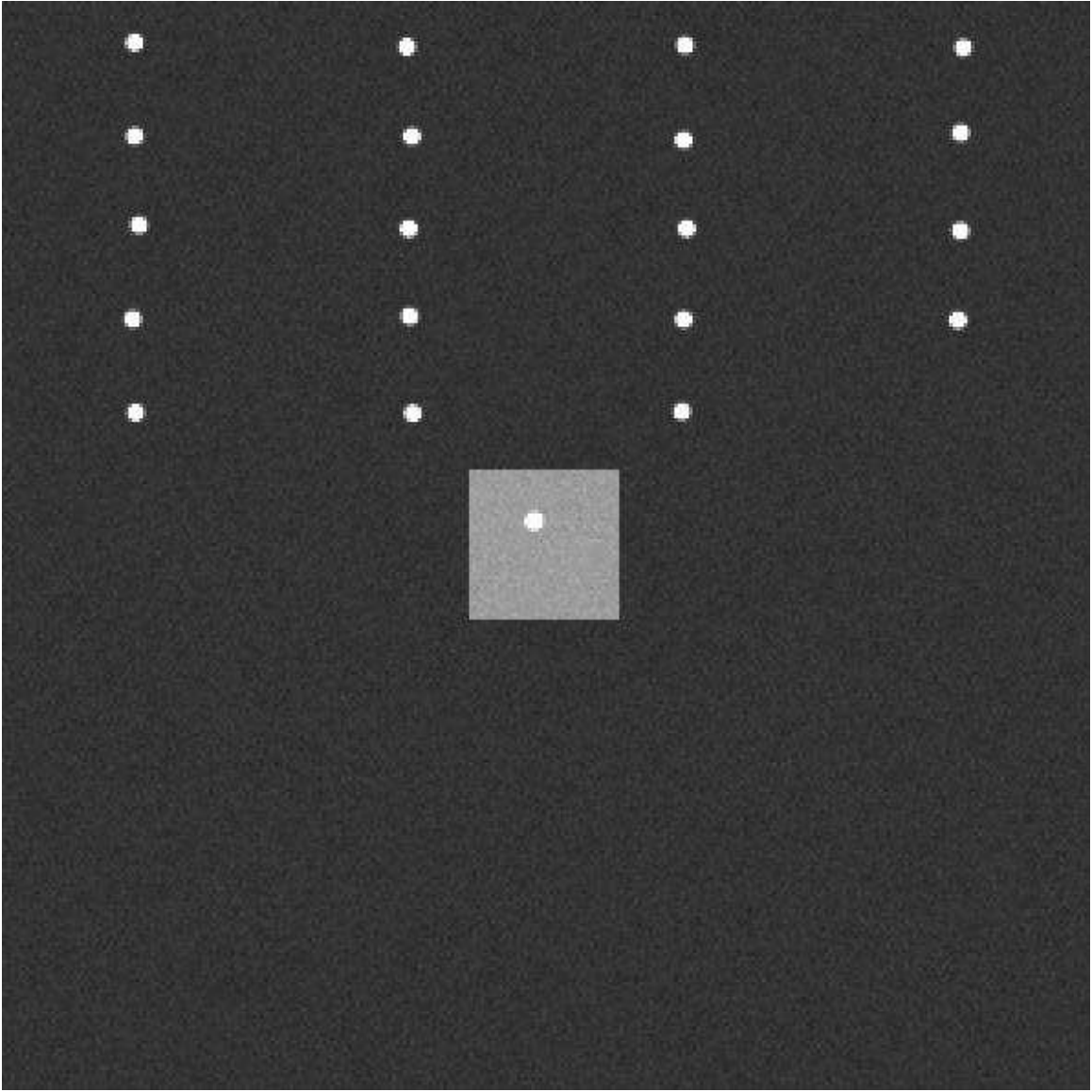}
 \caption{Simulated image with 19 reference stars in a 
          low background and a ``supernova''
          immersed in a ``square galaxy''.
          \label{fig:simflat} }
\end{figure}

To test this hypothesis, 
we created a set of simulated images
with properties similar to those acquired at RIT,
and subjected them to exactly the same measurement methods
as we used on our actual images.
We started with a simplified situation:
a set of reference stars of identical brightness
on a uniform background,
and a single ``supernova'' immersed in a ``square galaxy''
region of uniform higher intensity;
see 
Figure \ref{fig:simflat}.
The stars are modeled as gaussians of FWHM 3.0 pixels,
matching the typical seeing at RIT,
and the gain in the image is set to 2.2 electrons per count,
matching the properties of the SBIG ST-9E camera.
The brightness of the ``square galaxy'' is set to
620 counts, typical for the region near SN 2016coj in $R$-band images.
Note that the stars are placed at intervals with
small random variations,
ensuring that they appear at a wide range of sub-pixel locations.
We ran a number of instances of each simulation,
shifting both the reference stars and the ``supernova'' 
by small random sub-pixel positions each time.

\begin{figure}
 \plotone{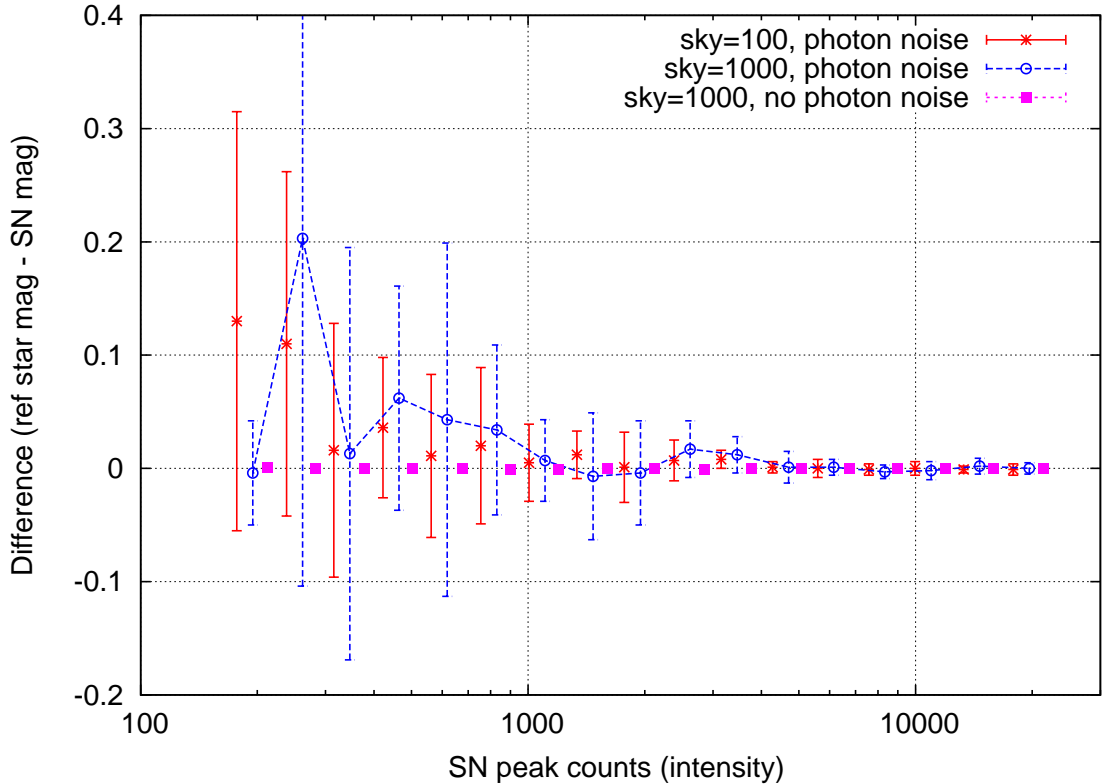}
 \caption{Results of photometry after rotation and subtraction
          of the simulated images with a ``square galaxy.''
          \label{fig:testrotsubb} }
\end{figure}

In this simplified situation,
we chose as the center of rotation the geometric center 
of the image.
After making a copy and rotating it around this center
by $180^\circ$,
we subtracted the copy from the original,
leaving both reference stars and ``supernova'' 
in a near-zero background;
but the pixels surrounding the supernova might be
noisier than those surrounding the reference stars.
As a sanity check that our software was not introducing
errors of its own,
we ran simulations in which no photon noise was added
to the images: the background value was some fixed
value in all pixels, and the model gaussian for each star
was similarly exact.
Over a series of trials,
the reference stars and ``supernova'' were all set
to the same input brightness, 
increasing gradually throughout the trials until
their centers reached a value of 30,000 counts
(similar to the limit of linearity on our camera).
We would expect the stars and ``supernova'' 
all to have exactly the same
magnitude in this noiseless simulation.
The magenta symbols in 
Figure \ref{fig:testrotsubb}
show that the difference between the average 
reference star magnitude, and the ``supernova'' magnitude,
is indeed zero
under these conditions.

However, 
when we add photon noise to our simulations,
we find some differences between the 
magnitude of the reference stars and the ``supernova.''
We ran simulations with two background levels,
100 and 1000 counts per pixel,
roughly bracketing the range of sky levels
in real RIT images 
(which varies due to clouds, haze, and the aspect of the Moon).
Consider first the red asterisk symbols, 
which show the results under low sky conditions:
when the SN and stars are bright,
there is no significant difference in their measured magnitudes.
But when the stars are faint, noise in the sky background and
in their own signal leads both to increased scatter
and to discrepancy between the average value of the reference
stars and that of the ``supernova;''
the SN tends to be measured as brighter than the reference stars.
The amplitude of the difference grows to roughly $0.1$ magnitude
by the time the stars are too faint to detect reliably.
The blue circular symbols,
corresponding to higher and so noisier background sky levels,
show the same trend, but at an increased amplitude.

We conclude that point sources simply immersed in a higher background
of light will suffer from a systematic bias
under our measurement procedure,
appearing brighter than they ought to be.
However, the situation of SN 2016coj is even worse:
not only is it in a region of higher background level than
the comparison stars, but it sits in a strong spatial gradient.
What effect will this additional complication have on our
measurements?

\begin{figure}
 \plotone{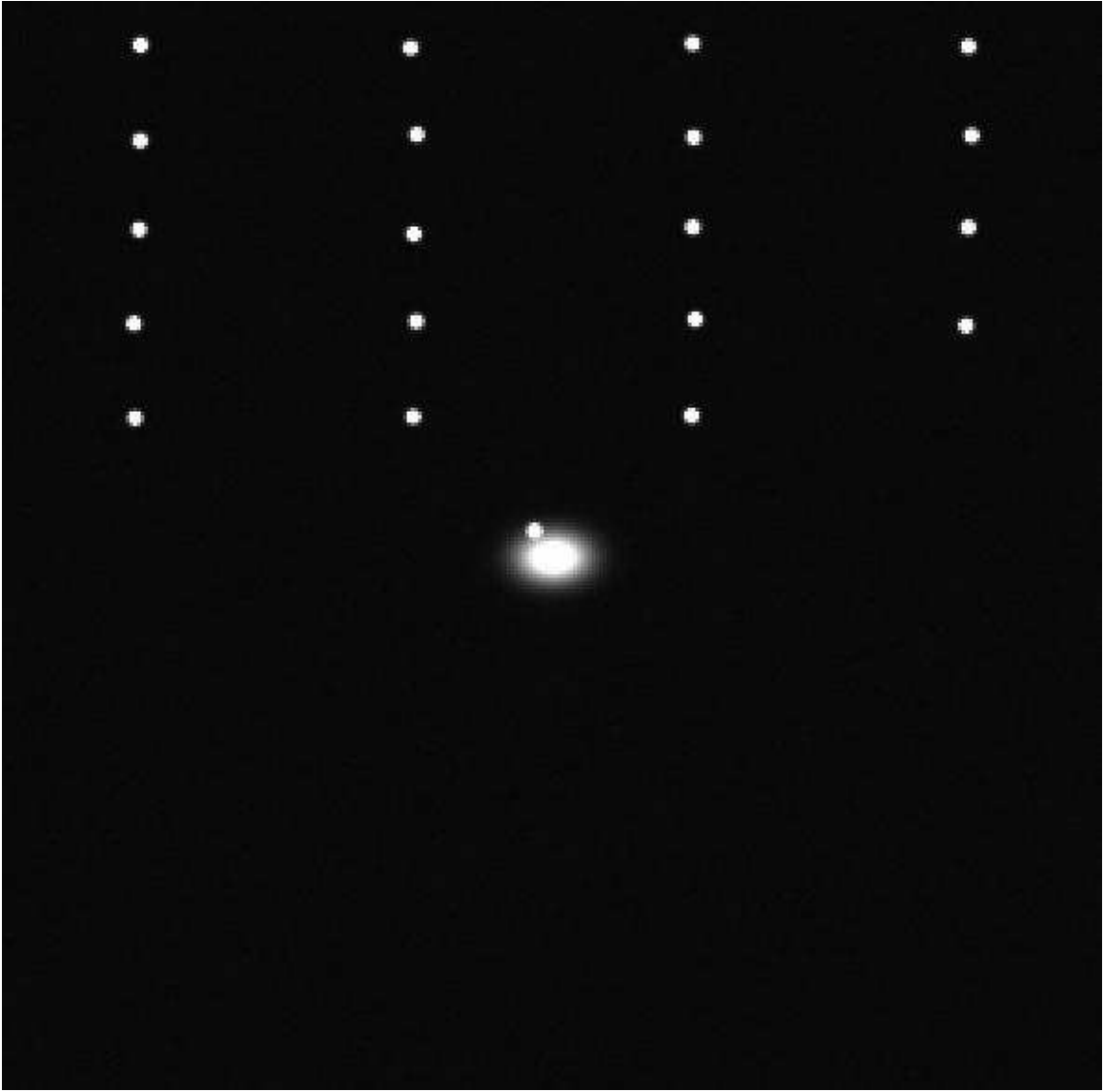}
 \caption{Simulated image with 19 reference stars in a 
          low background and a ``supernova''
          placed close to an artificial galaxy.
          \label{fig:simgalaxya} }
\end{figure}

We performed a very simple test by creating a toy model
of our real images.
We created an artificial galaxy by superposing a
central gaussian (FWHM = 3.0 pixels) 
and an extended and flattened component
(FWHM = 8.0 pixels, convolved with a kernel of FWHM = 6 along rows 
and FWHM = 12 along columns),
scaling the result so that it resembled the appearance of
NGC 4125 in our $R$-band images.
We then placed the ``supernova'' at an offset from the galaxy
similar to its actual offset.
Both the galaxy and the ``supernova'' were shifted in position by 
small random sub-pixel amounts in each realization of our simulations.
Figure \ref{fig:simgalaxya}
shows an example of these artificial images,
one in which the reference stars and SN are all
at the maximum brightness.

We then carried out a series of instances,
changing the brightness of the stellar objects
over a wide range;
at each level of brightness, we ran 10 realizations,
varying the positions of each source at the sub-pixel level
and generating different random values of photon noise.
For each realization, 
we carried out exactly the same measurement procedure
as we used for the real images:
\begin{itemize}
\item {a copy of the image was displayed on a computer screen}
\item {the user moved a cursor to the center of the galaxy, 
        pressed a key to initiate a calculation
        of the local centroid and display a radial profile,
        made adjustments to initial position 
        until satisfied that the center had
        been found correctly}
\item {a copy of the image was rotated around this position}
\item {the copy was subtracted from the original image}
\item {point sources in the residual image were automatically
        found and measured via aperture photometry}
\end{itemize}

\begin{figure}
 \plotone{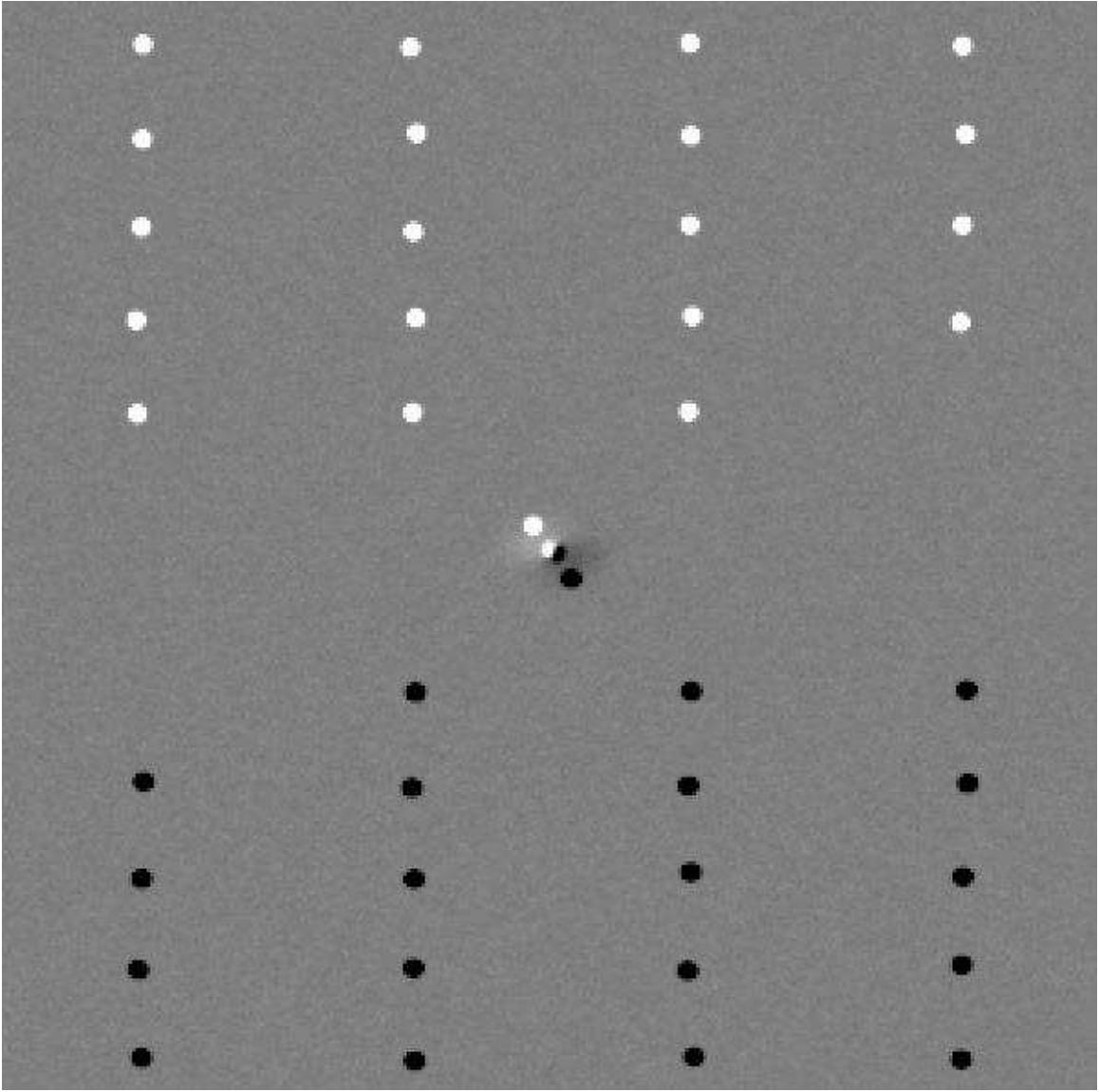}
 \caption{Simulated image with artificial galaxy
          after rotation and subtraction.
          \label{fig:simgalaxyb} }
\end{figure}

An example of one residual image is shown in
Figure \ref{fig:simgalaxyb}.
There is clearly imperfect subtraction of the galaxy's light
at its very center, as is seen in most of the real images
after this procedure.

\begin{figure}
 \plotone{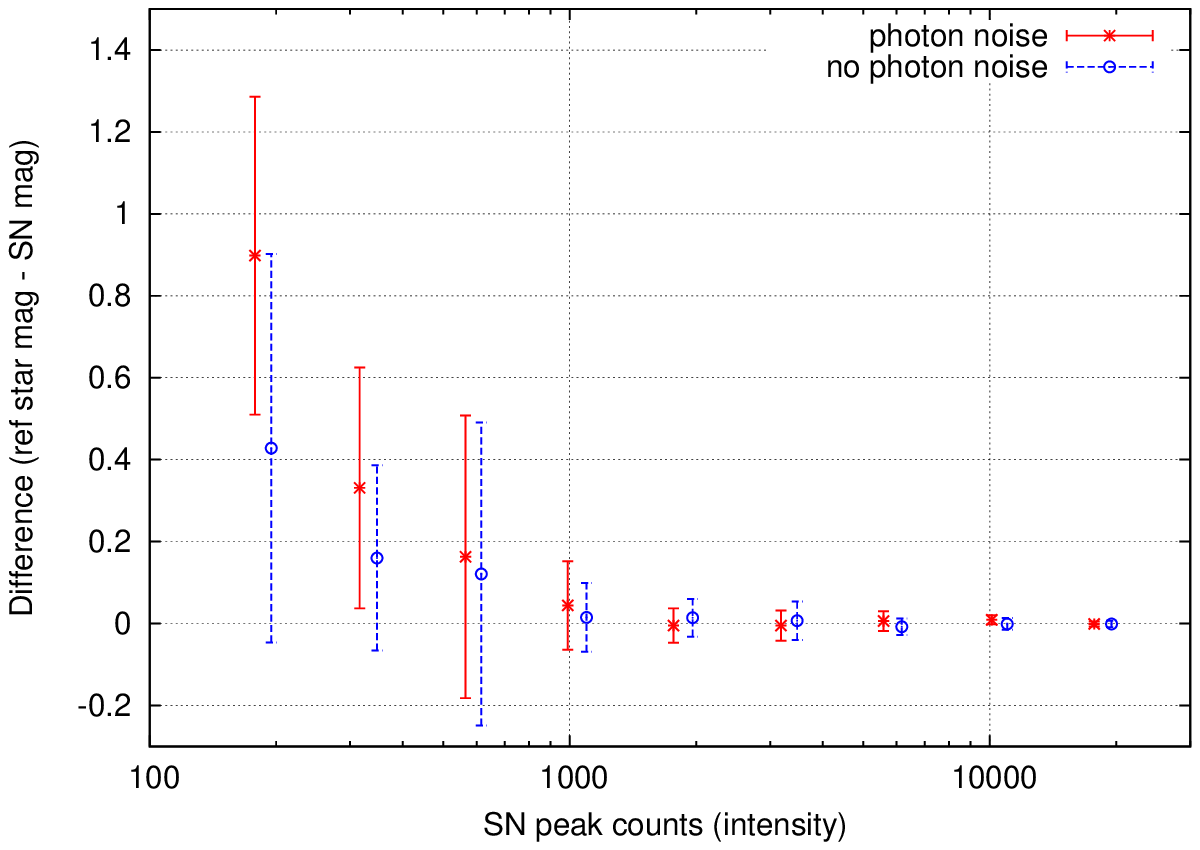}
 \caption{Results of photometry after rotation and subtraction
          of the simulated images with a realistic galaxy model.
          \label{fig:testrotsuba} }
\end{figure}

The results of photometry on these 
images are display in
Figure \ref{fig:testrotsuba}.
In this case, we fixed the brightness of the 
overall sky background to 100 counts for all realizations,
so it represents the optimistic end of the spectrum
of real conditions.
The general trend is similar to that in the 
``square galaxy'' simulations:
measurements of the SN appear brighter than those of the
reference stars, by an amount which increases as the SN fades.
However, there are important differences:
first, this systematic difference appears even in the
absence of photon noise;
this indicates that the software used to perform the
image analysis
(XVista \citealt{Tref1989})
is unable to compute the center of the galaxy accurately enough,
or perform the image rotation accurately enough, or both.
Second, 
note that the {\it amplitude} of the systematic difference
is much larger than in the ``square galaxy'' simulations:
the SN can appear about 1.0 magnitude brighter
it ought to be, 
instead of just 0.1 magnitude.

In order to reduce these systematic errors, one must improve the
method of subtracting the background contribution to the total
light within the photometric aperture.
Given the large FWHM of our images, we could not decrease the size
of the photometric aperture significantly;
given the location of the SN, close to the galaxy's nucleus,
increasing the size of the background annulus would make matters worse,
and, given the large FWHM, decreasing the size of the background annulus is not possible.
Rather than choosing some constant value per pixel for the
background contribution, 
one could do better by making a model of the galaxy's light
within the photometric aperture.
We will investigate this technique in the future;
it would require a substantial effort to modify the existing software.
The difficulty of proper background subtraction is, of course, 
the reason that many astronomers adopt 
the ``template subtraction'' method.

Now, in light of this information,
let us review the light curves shown in 
Figure \ref{fig:bvricurves}.
When the SN is bright, measurements from
RIT and NSO agree well;
but as the SN fades, an offset between the two datasets
appears, with the NSO measurements slightly fainter.
The offset is largest in the $I$-band
and smallest in the $B$-band;
we ascribe this trend with wavelength to the color
of the galaxy's light.
The starlight of NGC 4125 is more prominent at long wavelengths,
making the background at the location of SN 2016coj
brightest (relative to the SN) in the $I$-band.
Since the NSO data have both a smaller PSF
and a higher signal-to-noise ratio,
they suffer from less contamination by the galaxy's light.

At early times,
the SN was bright enough that any systematic 
bias was at most comparable to the random uncertainties
in each measurement;
but that is certainly not true for the late times.
We recommend that readers use with caution the latest
measurements presented here.
We suggest that greater weight be given to the NSO values
at late times;
it might be profitable to ``warp'' the RIT measurements at late
times to match the final NSO magnitudes.

\newpage%%%%%%%%%%%%%%%%%%%%%%%%%%%%%%%%%%%%%%%%%%%%%%%%%%%%%%

\end{document}